\documentclass[12pt,preprint]{aastex}
\usepackage{emulateapj5}
\begin{document}
 
\title{The Validity of the Adiabatic Contraction Approximation for Dark Matter Halos}
\author{Roland Jesseit, Thorsten Naab, and Andreas Burkert}
\affil{Max-Planck-Institut f\"{u}r Astronomie\\
 K\"onigstuhl 17, D-69117 Heidelberg, Germany}
\email{jesseit@mpia-hd.mpg.de, burkert@mpia-hd.mpg.de, naab@mpia-hd.mpg.de}
\shorttitle{Adiabatic Contraction of Dark Matter Halos}
 
\begin{abstract}
We use high resolution numerical simulations to investigate the adiabatic contraction
of dark matter halos with a Hernquist density profile. We test the response of the halos to
the growth of additional axisymmetric disk potentials with various central concentrations
and the spherically symmetric potential of a softened point mass. Adding the potentials 
on timescales that are long compared to the dynamical time scale of the halo, the contracted halos have
density profiles that are in excellent agreement with analytical predictions based on the conservation of
the adiabatic invariant $M(r)r$. This is surprising as this quantity  is
strictly conserved only for particles on circular orbits and in spherically symmetric potentials.
If the same potentials are added on timescales that are short compared to the dynamical
timescale, the result depends strongly on the adopted potential. The adiabatic approximation  
still works for disk potentials. It does, however, fail for the central potential.
\end{abstract} 


\section{Introduction}
The hierarchical clustering model is the paradigm for galaxy formation today. In this
cosmological scenario, structure forms around peaks of primordial dark matter density fluctuations. The 
baryonic matter, which can dissipate energy through radiation, cools and falls into the center
of its surrounding dark halo. 
The question of how a spherical mass distribution, e.g. a galactic bulge, will respond to the growth of mass in
its center has been adressed by \citet{bw84}, hereafter BW84. They devised a simple recipe for predicting the density
profile of a contracted spherical density distribution. This recipe was used by \citet{blum86} to examine
contracted dark halos, assuming that the baryonic disk forms in the center so slowly that the time for the 
increase of mass inside an orbit of a dark particle is long compared to its orbital period. 
In a slowly varying potential the action integral $j_i = \int p_i dq_i$ is a conserved property
of the particle orbit, called an adiabatic invariant \citep{BT87}. Here, j is the action, p and q are the
phase space coordinates of the dark particle.
As a first approximation, BW84 assumed a spherical density distribution with particles moving on 
circular orbits. In this case, the radial action integral simplifies to the conservation of angular momentum. 
With $L=mvr$ and the circular velocity $v^2(r)=GM(r)/r$ we get the adiabatic invariant $M(r)r$.\\
Given that we know the final baryonic matter distribution, e.g. an exponential disk-like profile $M_b(r)$
and the initial dark halo distribution $M_i(r)$, we can construct the final dark matter distribution
$M_f(r)$ according to 
\begin{equation}
\label{invariant}
r_f[M_b(r_f)+M_f(r)]=r M_i(r),
\end{equation}
where $r_f$ is the final radius of a dark particle. This approximation is strictly valid only if the initial
mass distribution is spherically symmetric. The mapping between initial and final radius is unique, because the 
dark particles are dissipationless and their circular orbits do not cross.
The adiabatic compression has been used widely in estimating rotation curves in semi-analytical galaxy models 
\citep{rydgun87,ryd88,ryd91,flores,mo98}, in investigating the origin of the Tully-Fisher relation \citep{rix}, in 
analysing the core structure of dark matter halos \citep{bosch,marche} and in the formation of gaseous disks in cosmological
N-body halos (M. Steinmetz 2001, private communication). The adiabatic contraction assumption has
been tested qualitatively in a low-resolution study by \citet{blum86}. 

It is surprising that equation (\ref{invariant}) should hold in realistic situations, 
where a non-spherical galactic disk is added to a halo of particles which move on eccentric or chaotic 
orbits \citep{valluri}.
In this paper we investigate the validity of equation (\ref{invariant}), using high resolution numerical simulations.
Adding a disk potential to an equilibrium N-body halo, we can study its reaction in the adiabatic limit, as well as for
an abrupt change of the potential. In \S 2 we describe the model runs for different disk parameters and contraction times.
In \S 3 we discuss the results and in \S 4 we draw our conclusions and present suggestions for future work. 
 
\section{The Simulations}
The N-body halo is set up according to the distribution function devised by \citet{her90,her93}. Its density 
distribution is $\rho(r) = M_h r_h/[2 \pi r (r+r_h)^3]$, where $M_h$ is the total mass and $r_h$ is 
the scale length. The Hernquist halo has the same $\rho \propto r^{-1}$ dependence in the center as the universal dark 
matter profile found by \citet{NFW}, however with a finite total mass as the density in the outermost regions
decreases as $\rho \propto r^{-4}$. The baryonic component is represented by analytic external potentials of a disk
and a central point mass. We used the potential of an exponentially thin disk according to \citet{dehnen98}.
Its strength depends on the ratio of the disk mass $M_d$ to the disk scale length $r_d$. 
Following \citet{NEF}, we started with a very large disk scale length and kept the disk mass constant throughout
the simulation. The disk scale length is contracted linearly with time, where the contraction rate is a free parameter.
Large contraction time scales compared to the dynamical time scales correspond to the adiabatic limit. In the limit of
zero contraction time the halo will go through a phase of violent relaxation.
The halo-disk system was allowed to relax after the contraction phase for several dynamical time scales.
We applied a massive and a low-mass disk model with 20 \% (MD) and 5 \% (LD) of the total dark halo mass,
and with a typical final scale length of 0.14 $r_h$. 

In order to test the dependence on the concentration of the potential we used a softened point mass potential following
\begin{equation}
\Phi_p(r)= \frac{G M_p \arctan(r/r_p)}{r_p}.
\end{equation}
\noindent Similar to the disk case we kept its mass $M_p$ constant and shrank the smoothing length to a final value of  
$r_p=0.03$ (test case PC). We also tested the case where $r_p$ was kept constant and $M_p$ was allowed to grow (PG). 
For all calculations we chose  $G=M_h=r_h=1$, where G is the gravitational constant.
Simulations were performed with halos represented by
N = $1 \times 10^4$, $8 \times 10^4$, $2 \times 10^5$ and $1 \times 10^6$  particles in order 
to test the dependence of the results on the numerical resolution. All model parameters are listed in Table \ref{tableone}. 
The gravitational softening length $\epsilon$ was chosen according to the criterium of
\citet{merritt}. We used a time step at least a hundred times smaller than the dynamical time scale we would expect at 
the half mass radius for each model. The timestep, adequate for the runs with the highest number of particles, was
not enlarged for lower particle number runs. We used a newly developed tree code WINE 
(Wetzstein et al., in preparation) in combination with special purpose hardware GRAPE-5 \citep{kawai} at the MPIA, 
Heidelberg. The refined force accuracy criterium of \citet{sw} guarantees that the absolute force error stays below 
the precision of the GRAPE hardware.    
\section{Results}
\begin{figure}
\plotone{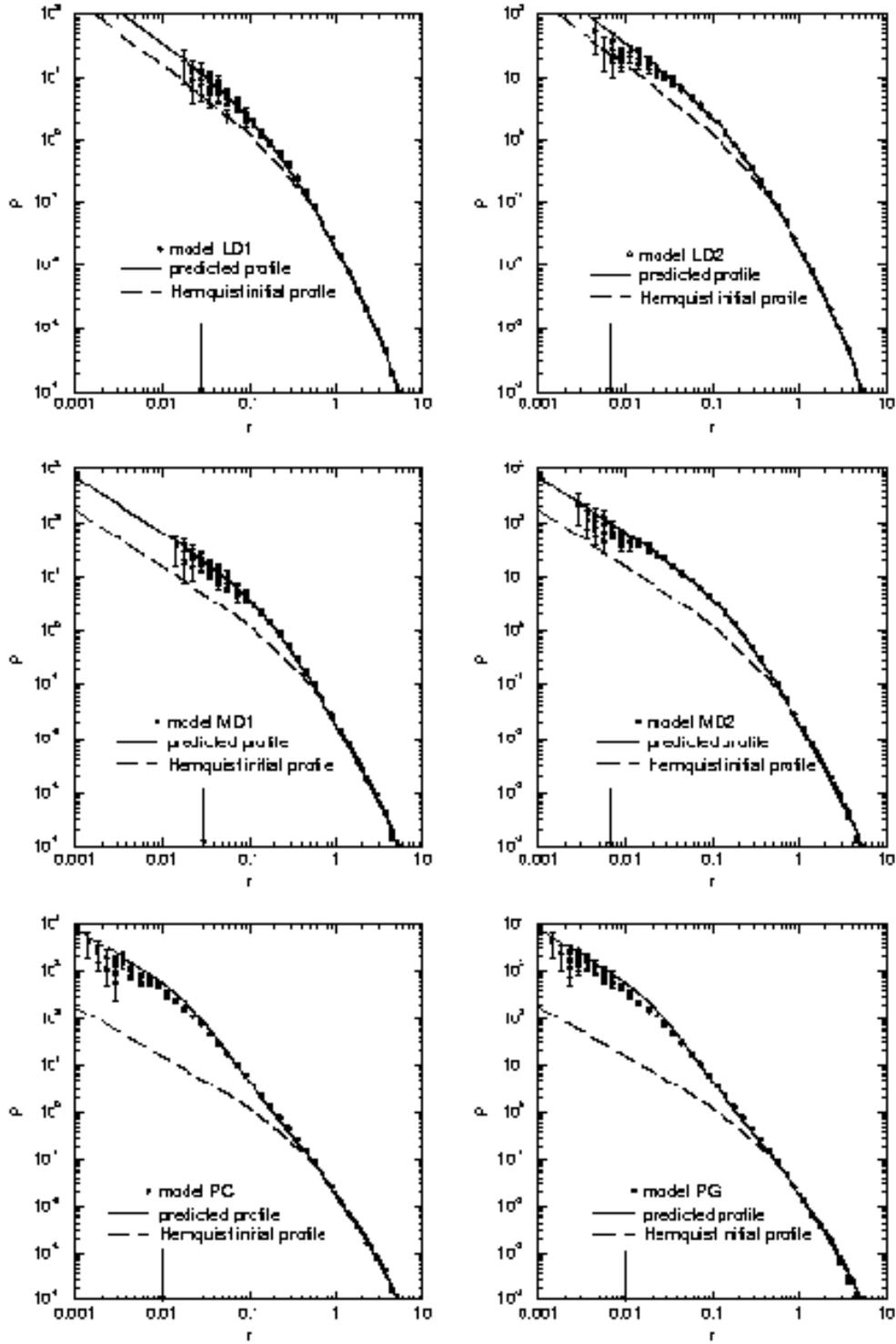}
\caption{Final density profiles of contracted dark halos. The upper two panels show the results for the models
	 LD1 and LD2, the following two for the models MD1 and MD2. The last two panels show the results for the point
	 mass. In the left panel the contracting scale length case and in the right panel the growing mass case is 
         shown. Arrows indicate the used softening length. \label{plotone}} 
\end{figure}

The four top panels in Fig. \ref{plotone} show the final density profiles of the contracted halos for two resolutions 
for the massive and the light disk, respectively (models LD1, LD2 and MD1, MD2). The error bars show the Poissonian
error. The softening length is indicated with an arrow on each plot. In the innermost parts of the halo, 
i.e. $r \leq 2 \epsilon$, the density is influenced by at least two effects: The softening length and fluctuations
in the density due to small particle numbers. To show the effect of the fluctuations we plot the density profile of 
four consecutive dumps taken shortly after the contraction has been completed. This is important as the fluctuations
sometimes exceed the 1 $\sigma$ Poissonian error. At radii larger than two softening lengths the analytical approximation
of BW84 gives a very good account of the matter distribution of the halo for every disk model. The two panels at the bottom
of Fig. \ref{plotone} show the point mass case which is also in very good agreement with the theoretical profile, 
independent of the way we grow the external potential. However, the density is somewhat lower than theoretically 
predicted inside the scale radius $r_p$.

\begin{figure}
\plotone{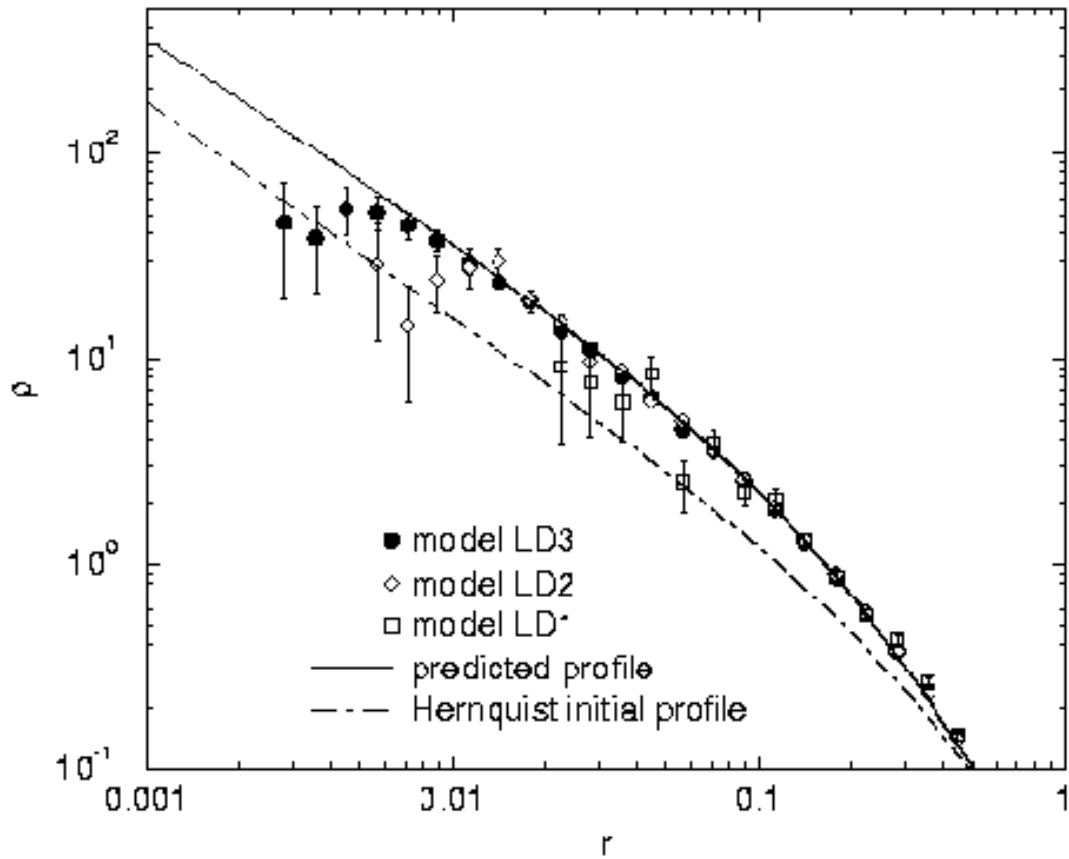}
\caption{Final density profiles of LD1,LD2 and LD3 are shown to examine the effects of numerical resolution.
	 \label{plottwo}} 
\end{figure}

Fig. \ref{plottwo} illustrates for the LD case that the agreement between numerical simulation and theoretical
profile improves with increasing particle number, as we are able to probe deeper into the center of the dark matter halo.

\begin{figure}
\plotone{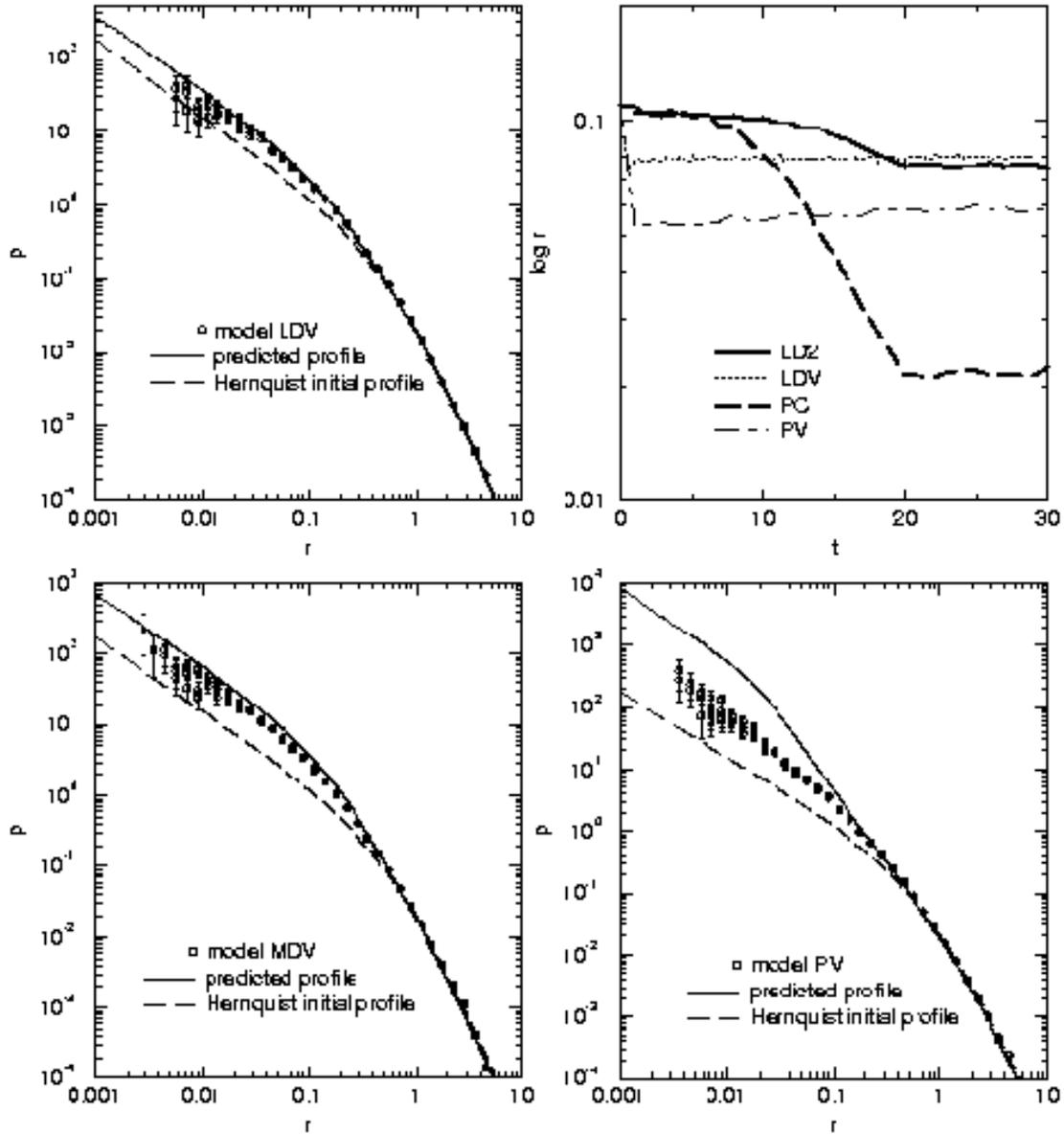}
\caption{Final density profiles after abrupt addition of external potential.  
        LDV top left, MDV bottom left and PV bottom right. In the top right panel a comparison of the evolution of 
	the 1\% mass shell radii between the adiabatic and the violent case of the models LD and P is shown. 
        \label{plotthree}} 
\end{figure}

In Fig. \ref{plotthree} we examine the response of the halo to an abrupt addition of the external potentials
(models MDV, LDV and PV). The profile with the light disk (top left) still matches the theoretical curve. Even in 
the case of the massive disk (bottom left), the predicted density distribution agrees well with the numerical model 
although we find a somewhat lower density inside a radius of one disk scale length. Adding the point mass potential 
instantaneously (bottom right) leads to a density distribution that deviates strongly from the
adiabatic prediction. The top right panel in Fig. \ref{plotthree} shows the time evolution of the 1\% 
mass shell radius in each case. For the LD models the mass shells stabilize at more or less the
same radius, though the detailed path of contraction is very different for LD2 and LDV.
For the more centrally concentrated potential (PV) this is not the case. It stabilizes at radii further outside
than in the adiabatic case. Consequently these density profiles are less concentrated in the center
than in the adiabatic case.
\section{Conclusions}
We find that the adiabatic approximation is very robust in the parameter space that is occupied by normal spiral or 
dwarf galaxies. This remains valid even in cases where the contraction timescales are short compared with the dynamical
time scales. Only for a point mass potential, a significant difference occurs. We conclude that the validity of the 
adiabatic approximation depends mainly on the concentration of the added potential. A better understanding of the 
underlying physics will require a detailed analysis of the orbital distribution of the particles prior and after the 
contraction phase. We will investigate this question in a subsequent paper. 
   
\acknowledgments
A program to calculate disk potentials was kindly provided by Walter Dehnen. 
We want to thank Hans-Walter Rix, Matthias Steinmetz and Chris Gottbrath 
for helpful comments and discussions.

\clearpage
 
\begin{deluxetable}{ccrrcc}
\footnotesize
\tablecaption{Model runs \label{tableone}}
\tablewidth{0pt}
\tablehead{
\colhead{Model} & \colhead{Mass/scale length}   & \colhead{N}   & \colhead{$\epsilon$} & 
\colhead{dt $[10^{-3}]$}  & \colhead{type} } 
\startdata
LD1 &0.05/0.14 & $1 \times 10^4$  &0.03 &6.25&adiabatic\\
LD2 &0.05/0.14 &$2 \times 10^5$ &0.007&6.25 &adiabatic\\
LD3 &0.05/0.14 &$1 \times 10^6$ &0.0035&6.25&adiabatic\\
LDV &0.05/0.14 &$2 \times 10^5$ &0.007&6.25 &violent\\
MD1 &0.2/0.14  &$1 \times 10^4$  &0.03 &1.57 &adiabatic\\
MD2 &0.2/0.14  &$2 \times 10^5$ &0.007&1.57 &adiabatic\\
MDV &0.2/0.14  &$2 \times 10^5$ &0.007&1.57 &violent\\
PG  &0.2/0.03   &$8 \times 10^4$  &0.01 &0.78 &point grow\\
PC  &0.2/0.03   &$8 \times 10^4$ &0.01 &0.78 &point contract\\
PV  &0.2/0.03   &$8 \times 10^4$  &0.01 &0.78 &point violent\\
\enddata

\tablecomments{Abbreviations for the model runs are LD = Light Disk, MD = Massive Disk, P = Point Mass,
 V = Violent, G = Mass Growth, C = Softening Length Contraction. Model parameters are N = number of particles 
and dt = integration time step. Second column shows mass to scale length ratios for point mass and disk models.}
\end{deluxetable}

\end{document}